# OUT PERFORMANCE OF CUCKOO SEARCH ALGORITHM AMONG NATURE INSPIRED ALGORITHMS IN PLANAR ANTENNA ARRAYS

A.Sai Charan[1], N.K.Manasa[2], Prof. N.V.S.N. Sarma[3]

[1, 2, 3] Department Of Electronics and Communication Engineering,

National Institute of Technology, Warangal

[1]charanadd@gmail.com
[2]manasank1992@gmail.com
[3]sarma@nitw.ac.in

## ABSTRACT

*In this modern era a great deal of metamorphism is observed around us which eventuate due to some minute modifications and innovations in the area of Science and Technology. This paper deals with the application of a meta heuristic optimization algorithm namely the Cuckoo Search Algorithm in the design of an optimized planar antenna array which ensures high gain ,directivity, suppression of side lobes, increased efficiency and improves other antenna parameters as well[1], [2] and [3].*

## KEYWORDS

*Meta-Heuristic, Side Lobe Suppression, Gain, Directivity, Side Lobe Level (SLL).*

## 1. INTRODUCTION

Antenna optimization techniques have made a breakthrough in the Communication domain. They have contributed vividly to modern wireless communications in the form of smart antennas which are antenna arrays that adjust their own beam pattern to accentuate signals of interest and concurrently reducing the radio frequency interference. In the field of antennas, Cuckoo Search Algorithm (CSA) was first applied for side lobe suppression in linear antenna array by distance modulation.

Large arrays are complex to build, have increased fabrication and set up cost and are heavier at the same time. Therefore reducing antenna element weight from the array is desirable without degrading the performance of the array. But here we are not reducing the mass of the antenna array elements, only the weight of the antenna elements(current)are adjusted in order to achieve minimum side lobe level.

We opt a technique based on density tapering to lower side- lobes in the array by monotonically decreasing the magnitude of weights away from the centre of the array.

## 2. REVIEW OF VARIOUS TECHNIQUES

Owing to high adaptability and ability to optimize multi-dimensional problems, several evolutionary algorithms have been proposed such as Particle Swarm Optimization (PSO), Invasive Weed Optimization (IWO), Genetic Algorithm (GA), etc. These algorithms are associated with following drawbacks which make them unreliable.

i) The PSO could not work out the problem of scattering and optimization.

ii) The IWO require the genes of minimum one parent species to be forwarded to next generation and

iii) The GA has a poor fitness function which generates bad chromosome blocks in spite of the fact that only good chromosome blocks cross over. Also no assurance is given whether the GA will find a global optimum solution [4].

This paper has explored a choice of antenna array synthesis, the (CSA) [5], to overcome the above mentioned problems and to yield promising results.

## 3. PLANAR ARRAYS

Planar array is a two dimensional configuration of elements arranged to lie in a plane. The planar array may be thought of as an array of linear arrays. The elements are arranged in a matrix form having a phase shifter. The planar arrangement of all antenna elements forms the complete phased array antenna. There are wide spread applications of planar antenna arrays which involve the suppression of side lobes. The signals radiated by individual antennas determine the effective radiation pattern of the array. They are used to point a fixed radiation pattern or to scan a region rapidly in the azimuthal plane. Several methods have been developed for the design of planar antenna array but all those methods pertain to other nature inspired optimization algorithms.

Planar antenna array optimization has been implemented earlier using Fuzzy GA [6]. Direction angle (reference angle) is considered with the plane of planar antenna array. This paper deals with the design of a planar antenna array by using CSA.

## 4. CUCKOO SEARCH ALGORITHM (CSA)

CSA is one of the modern nature inspired meta-heuristic algorithms. The Greek terms "meta" and "heuristic" refer to "change" and "discovery oriented by trial and error" respectively. Various techniques are used to minimise the constraints associated with the problem in order to obtain a global optimum solution.

Cuckoos are attractive birds. The attractiveness is owing to the beautiful sounds produced by them and also due to their reproduction approach which proves to be combative in nature. These birds are referred to as brood parasites as they lay their eggs in communal nests. They remove the eggs in the host bird nest in order to increase the hatching probability of their own eggs.

There are three types of brood parasites - the intraspecific brood parasite, cooperation breed and nest take over type. The host bird involves in direct combat with the encroaching cuckoo bird. If the host bird discovers the presence of an alien egg, it either throws away the egg or deserts the nest. Some birds are so specialized that they have the characteristic of mimicking the colour and the pattern of the egg which reduces the chances of the egg being left out thereby increasing their productivity [7].

The timely sense of egg laying of cuckoo is quite interesting. Parasitic cuckoo birds are in search of host bird nests which have just laid their own eggs. In general the cuckoo birds lay their eggs earlier than the host bird's eggs in order to create space for their own eggs and also to ensure that a large part of the host bird feed is received by their chicks.

## 5. PRINCIPLE BEHIND CUCKOO SEARCH ALGORITHM

Each cuckoo bird lays a single egg at a time which is discarded into a randomly chosen nest. The optimum nest with great quality eggs is carried over to next generations. The number of host nests is static and a host can find an alien egg with a probability (Pa) [0, 1], whose presence leads to either throwing away of the egg or abandoning the nest by the host bird [8].

One has to note that each egg in a nest represents a solution and a cuckoo egg represents a new solution where the objective is to replace the weaker fitness solution by a new solution.

*The flowchart for CSA is as shown which involves the following steps:*

*Step (1) - Introduce a random population of n host nests, $X_i$.*

*Step (2) - Obtain a cuckoo randomly by Levy flight behaviour, i.*

*Step (3) - Calculate its fitness function, $F_i$.*

*Step (4) - Select a nest randomly among the host nests say j and calculate its fitness, $F_j$.*

*Step (5) - If $F_i < F_j$, then replace j by new solution else let j be the solution.*

*Step (6) - Leave a fraction of Pa of the worst nest by building new ones at new locations using Levy flights.*

*Step (7) - Keep the current optimum nest, Go to Step (2) if T (Current Iteration) < MI (Maximum Iteration).*

*Step (8) - Find the optimum solution.*

Important Stages involved in CSA are:

*i) Initialization:* Introduce a random population of n host nest ($X_i = 1, 2, 3...n$).

*ii) Levy Flight Behaviour:* Obtain a cuckoo by Levy flight behaviour equation which is defined as follows:

$$X_i(t+1) = X_i(t) + \alpha \oplus \text{Levy}(\lambda), \alpha > 0 \quad (1)$$

$$\text{Levy}(\lambda) = t^{(-\lambda)}, 1 < \lambda < 3 \quad (2)$$

*iii) Fitness Calculation:* Calculate the fitness using the fit- ness function in order to obtain an optimum solution. Select a random nest, let us say j. Then the fitness of the cuckoo egg (new solution) is compared with the fitness of the host eggs (solutions) present in the nest. If the value of the fitness function of the cuckoo egg is less than or equal to the fitness function value of the randomly chosen nest then the randomly chosen nest (j) is replaced by the new solution.

$$\text{Fitness Function} = \text{Current Best Solution} - \text{Previous Best Solution} \quad (3)$$

Since the Fitness function = Current best solution - Previous best solution, the value of the fitness function approaching the value zero means that the deviation between solutions decreases due to increase in the number of iterations.

The conclusion is that if the cuckoo egg is similar to a normal egg it is hard for the host bird to differentiate between the eggs. The fitness is difference in solutions [10] and the new solution is replaced by the randomly chosen nest. Otherwise when the fitness of the cuckoo egg is greater than the randomly chosen nest, the host bird recognizes the alien egg, as a result of which it may throw the egg or forsake the nest.

*The various stages involved in the working of this algorithm are explained in the flow chart:*

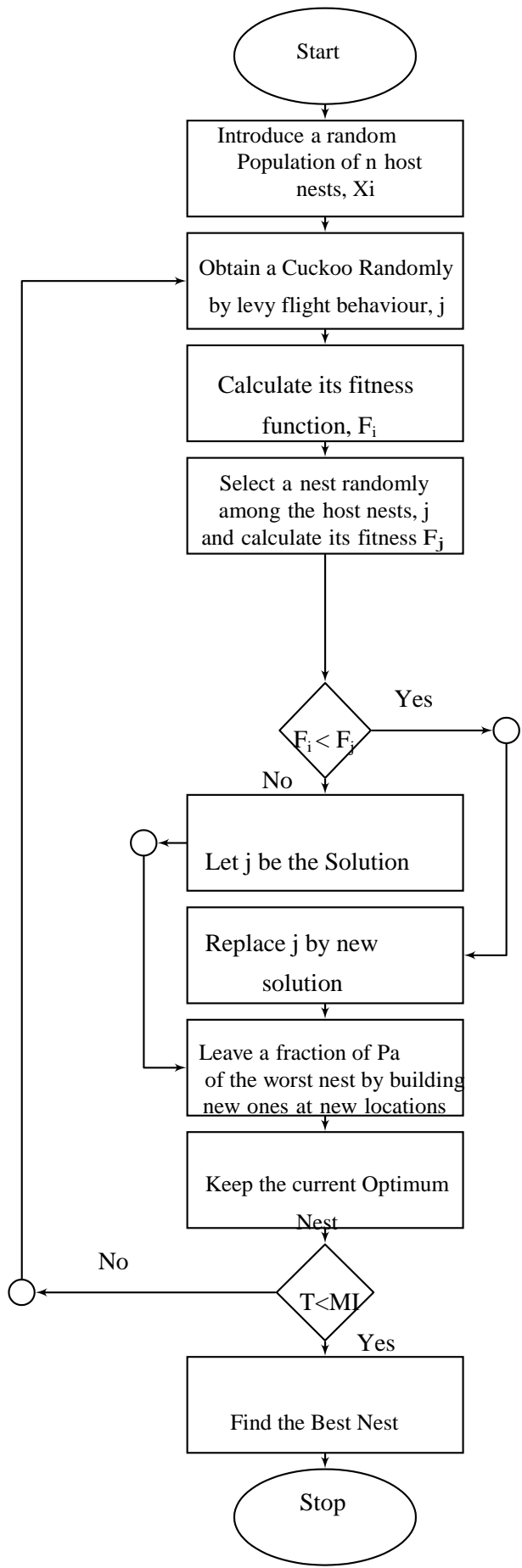

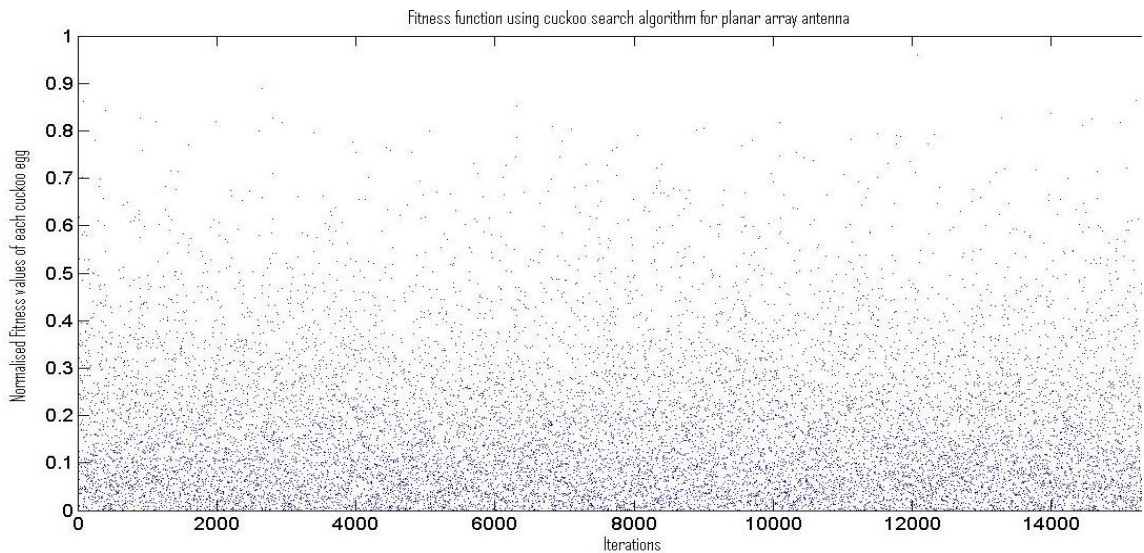

Figure 1. Fitness function values for planar antenna array of 18x18 elements

From the fitness function graph it can be observed that as the number of iterations increases, the value of the fitness function graph approaches to zero.

*iv) Termination:* In the current iteration the solution is compared and the best solution is only passed further which is done by the fitness function. If the number of iterations is less than the maximum then it keeps the best nest.

After the execution of the initialisation process, the levy flight and the fitness calculation processes, all cuckoo birds are prepared for their next actions. The CSA will terminate after maximum iterations [MI], have been reached.

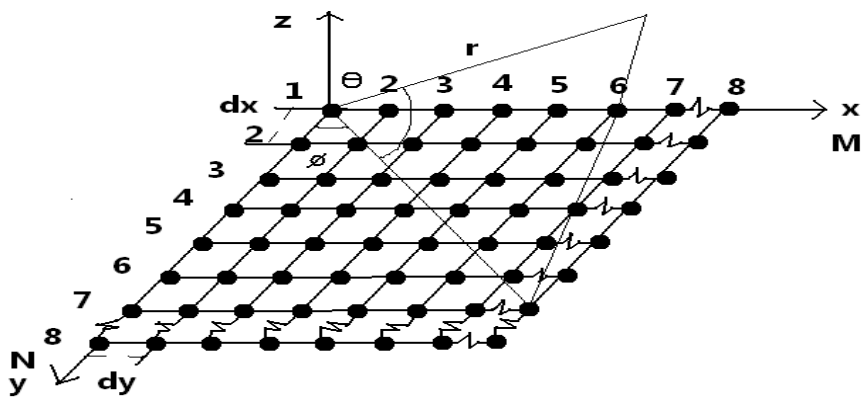

Figure 2. Planar Antenna Array Set–Up

## 6. TECHNICAL DETAILS

### 6.1. Synthesis of Planar Antenna Array

Consider a planar antenna array which consists of M-by-N rectangular antennas which are spaced equally [10]. They have been arranged in a regular rectangular array in the x-y plane. The inter-element spatial arrangement is

$$d = dx = dy = \lambda/2 = R_0$$

Where λ is the wavelength
The outputs are summed up in order to provide a distinct output.

$$F_s(\theta, \varphi) = \frac{f(\theta,\varphi)}{F_{s\,msx}} \sum_{m=1}^{M} I_m\, e^{j((m-1)k_x dx)+\psi_m} \sum_{n=1}^{N} I_n\, e^{j((n-1)k_y dy)+\psi_n}$$

Where $k_x = \frac{2\pi}{\lambda} \sin\theta \cos\varphi$, $k_y = \frac{2\pi}{\lambda} \sin\theta \sin\varphi$

## 6.2 Number of Cuckoo Birds

This parameter decides number of Cuckoo birds being initialized in the field space.

## 6.3. Step Size

In case of CSA, step size refers to the distance covered by a cuckoo bird for a fixed number of iterations. It is preferred to have an intermediate step size in order to obtain an effective solution. If the step size is too large or too small it leads to deviation from the required optimum solutions [7].

## 7. FIGURES AND TABLES

Table 1: SLL values for various sizes of Planar Antenna Array

| M | N | $P_a$ | SLL (in dB) | Main Lobe Range | Φ(in degrees) | Figure No. |
|---|---|---|---|---|---|---|
| 11 | 11 | .25 | -28.8 | [79.8, 100.2] | 90 | 4 |
| 16 | 16 | .25 | -29.2 | [83, 97] | 0 | 6 |
| 20 | 20 | .25 | -32.3 | [84.2, 95.8] | 90 | 7 |

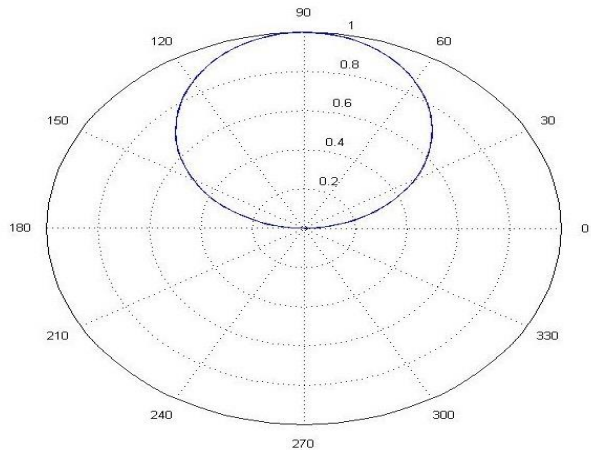

Figure 3.  Polar pattern of a single antenna array element

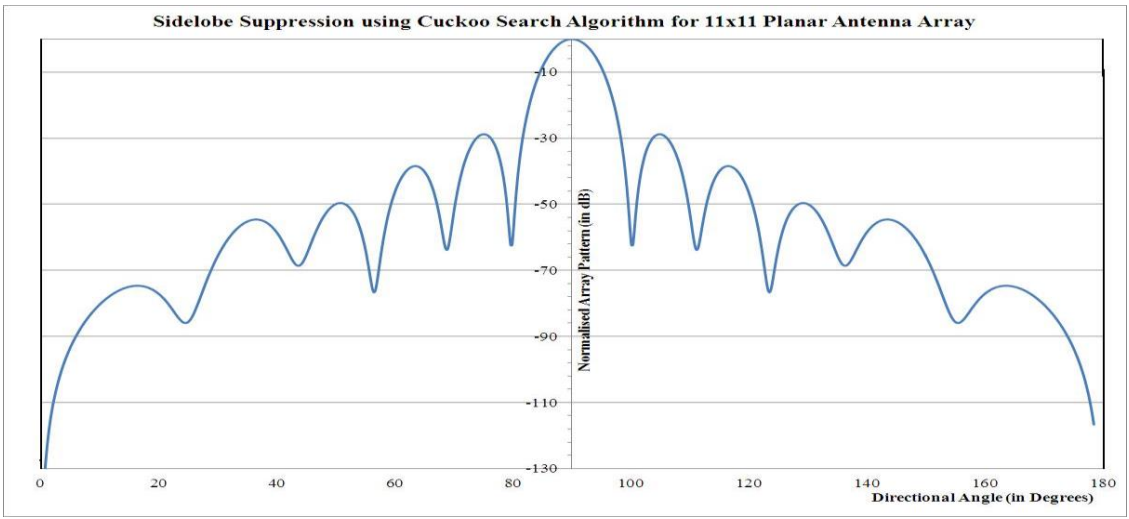

Figure 4. Radiation Pattern for a planar antenna array of 121 elements and φ= 0 degrees

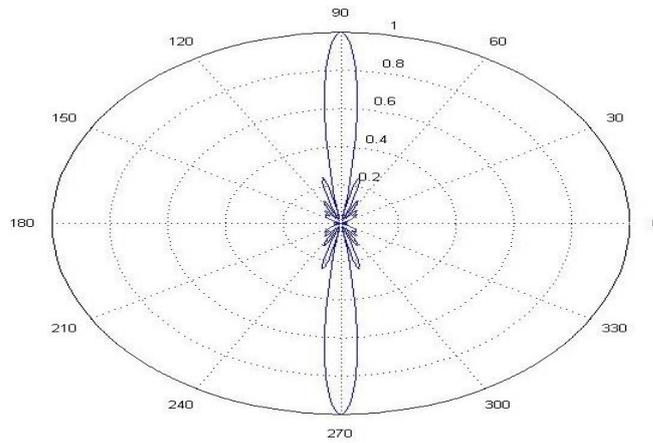

Figure 5. Polar pattern for the radiation of an 11X11 planar antenna array

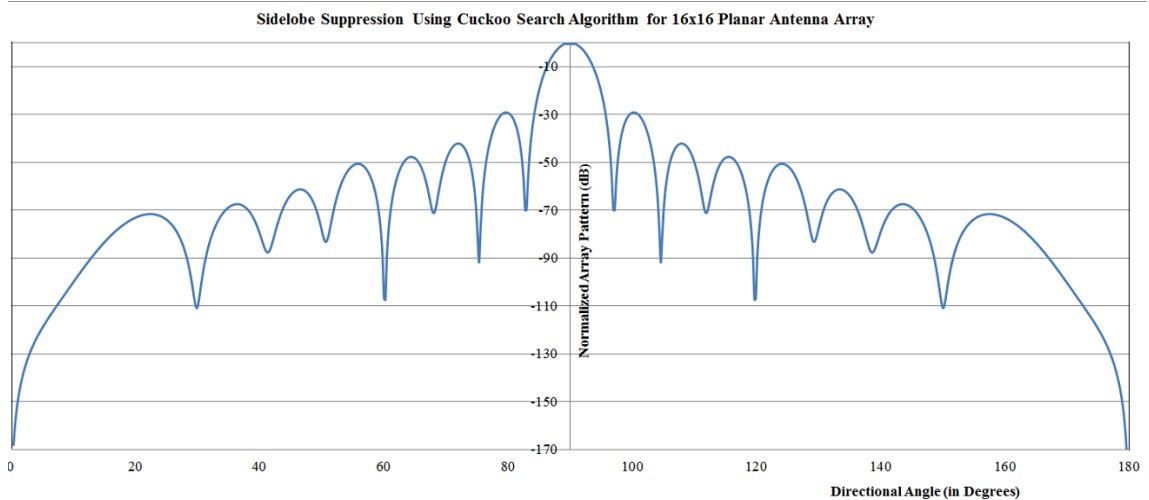

Figure 6. Radiation Pattern for a planar antenna array of 256 elements and φ= 0 degrees

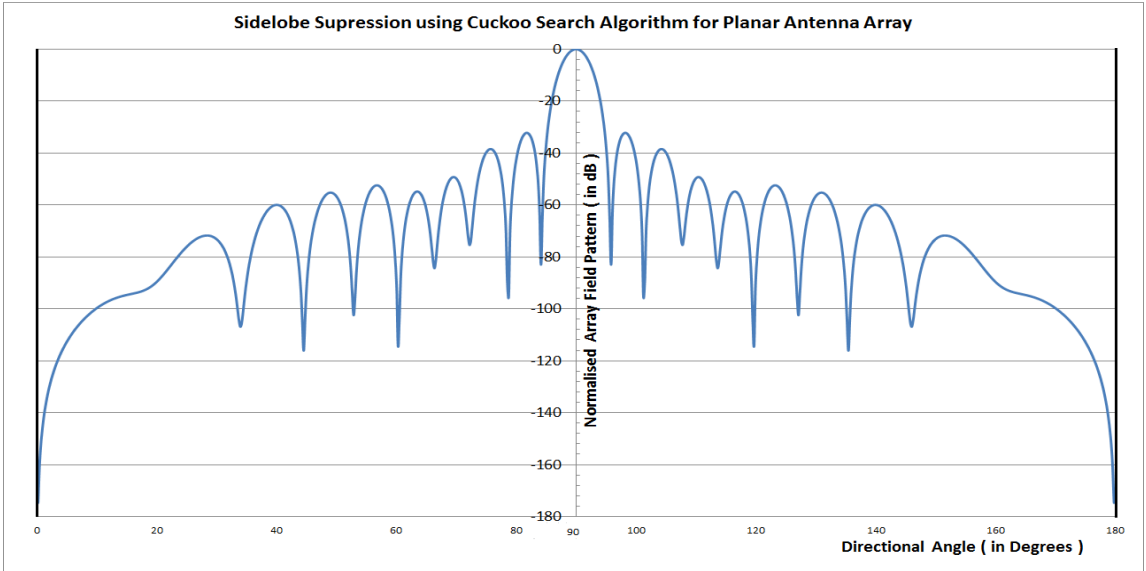

Figure 7. Radiation Pattern for a planar antenna array of 400 elements and φ = 90 degrees

## 8. Comparing CSA with GA and PSO

Consider the results for GA from the reference papers [2] and [13]. From paper [2], for N=16 elements -16.07 dB of SLL was achieved. From paper [13], for N=16 elements -28.47 dB of SLL was achieved.

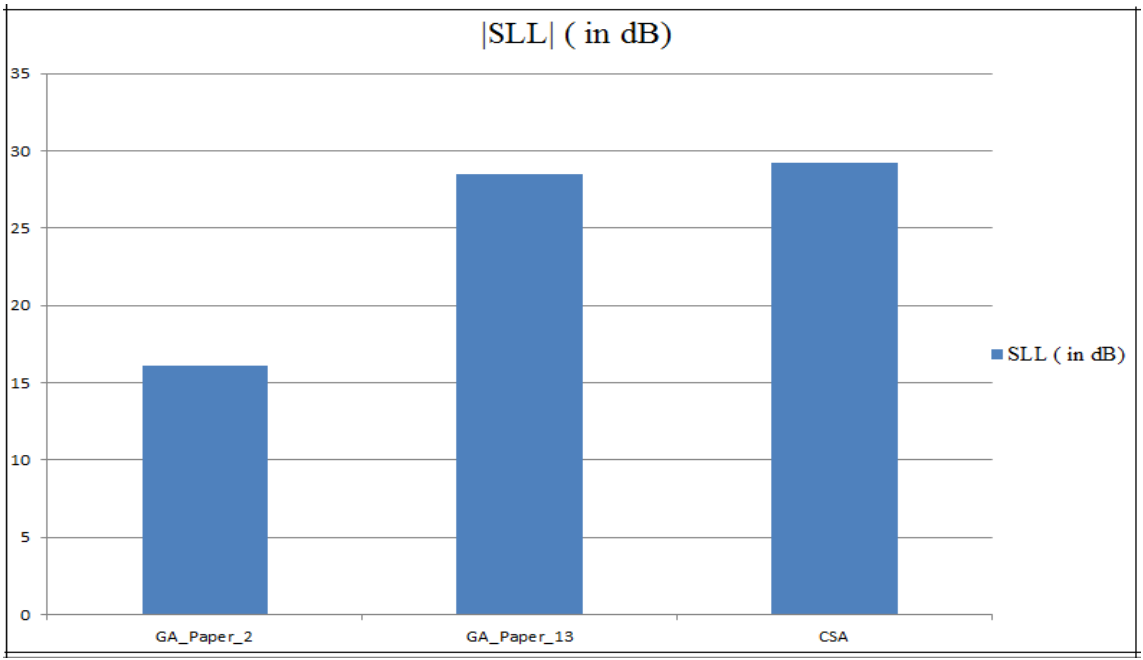

Figure 8. Comparison of CSA with GA using reference papers [2] and [13]

Consider the results for PSO from the reference paper [14]. From paper [14], for N=20 elements -16.2037 dB of SLL was achieved.

From paper [2], for N=20 elements -21.05 dB of SLL was achieved. From paper [13], for N=20 elements -28.59 dB of SLL was achieved.

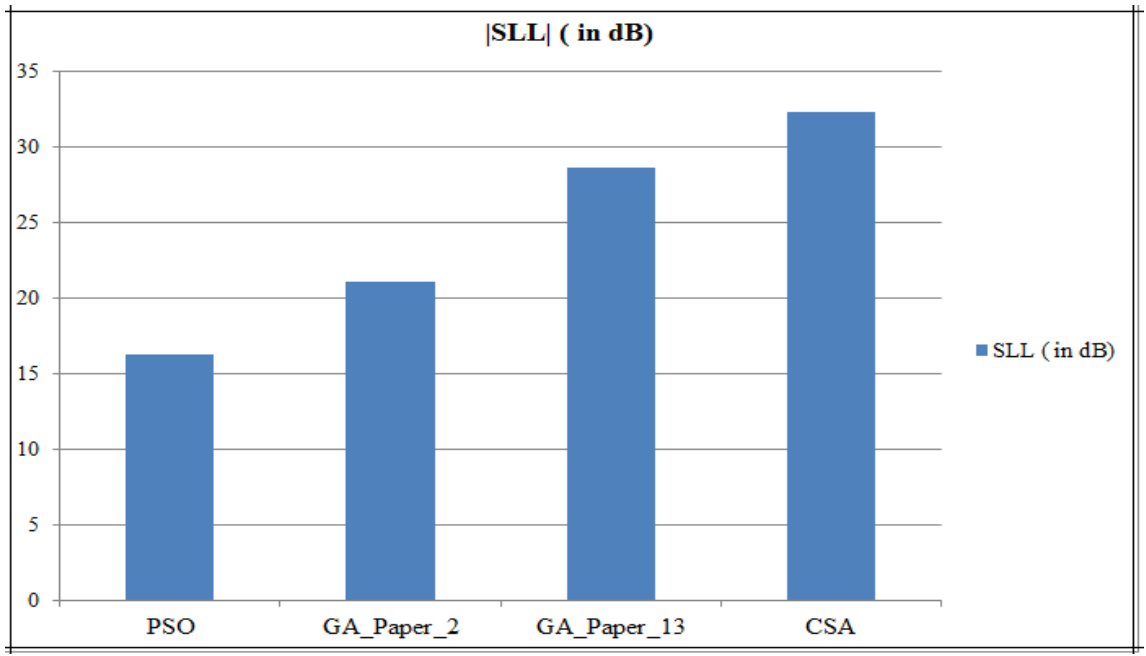

Figure 9. Comparison of CSA with GA and PSO using reference papers [2], [13] and [14]

## 9. OBSERVATIONS

When φ = 0 degrees, Maximum iterations = 500, a narrow beam is obtained as the best optimum solution for a large value of the number of iterations (optimum value) [11]. When maximum iteration is 150, main lobe appears to be spread over a wide range of direction angle (θ). For increase in the number of antenna elements in a planar antenna array a narrow beam is achieved correspondingly. The same field pattern is obtained for φ=π/2 degrees, maximum iterations of 500.

CSA in Planar Antenna Arrays is compared with GA and PSO that were implemented in Linear Antenna Arrays which can be used as sufficient data to come to a conclusion about the out performance of CSA compared to other Nature Inspired Algorithms.

The directivity of an isotropic antenna is unity as power is radiated equally in all directions [Fig 3]. In case of other sources such as omnidirectional antennas, sectoral antennas, directivity is greater than unity. Directivity can be considered as the figure of merit of directionality as it is an indication of the directional properties of the antenna with respect to an isotropic source. This shows that for any alignment of planar antenna array the same field pattern will be obtained which promotes beam steering in RADAR applications.

## 10. CONCLUSIONS

CSA is very easily applicable among all the nature inspired meta-heuristic algorithms since it provides the optimum solution. The implementation of CSA led to a tremendous increase in directivity [Fig 4] which promotes long distance communication. Gain of Antenna Array is also increased by using CSA.

## ACKNOWLEDGEMENTS

The authors are very thankful to Prof. N.V.S.N. Sarma for guiding them throughout the project.

## REFERENCES

[1] Ehsan Valian, Shahram Mohanna, Saeed Tavakoli, Improved Cuckoo Search Algorithm for Feed Forward Neural Network Training, International Journal for Artificial Intelligence and Applications (IJAIA) ,Vol.2, No.3, Pp. 36-43, July 2011.

[2] Pallavi Joshi, Nitin Jain, Optimization of Linear Antenna Array Using Genetic Algorithm for Reduction in Side Lobe Level and to Improve Directivity, International Journal of Latest Trends in Engineering and Technology(IJLTET), Vol.2, Issue 3, Pp. 185-191, May 2013.

[3] Khairul Najmy ABDUL RANI, Mohd.Fareq ABD MALEK, Neoh SIEW-CHIN, Nature Inspired Cuckoo Search Algorithm for Side Lobe Suppression in a Symmetric Linear Antenna Array, RADIO ENGINEER-ING, Vol.21, No.3, Pp. 865-974, September 2012.

[4] Ch.Ramesh, P.Mallikarjuna Rao, Antenna Array Synthesis for Suppressed Side Lobe Level Using Evolutionary Algorithms, International Journal of Engineering and Innovative Technology(IJEIT), Volume 2, Issue 3, Pp. 235-239, September 2012.

[5] Ehsan Valian, Shahram Mohanna, Saeed Tavakoli, Improved Cuckoo Search Algorithm for Global Optimization, International Journal of Communications and Information Technology, IJCIT-2011-Vol.1-No.1, Pp. 31-44, Dec 2011.

[6] Boufeldja Kadri, Miloud Boussahla, Fethi Tarik Bendimerad, Phase-Only Planar Antenna Array Synthesis with Fuzzy Genetic Algorithms, IJCSI, International Journal of Computer Science Issues, Vol.7, Issue 1, No.2, Pp. 72-77, January 2010.

[7] Xin-She-Yang, Nature Inspired Meta-heuristic Algorithm, 2nd Edition, Luniver Press, 2010.

[8] Monica Sood, Gurline Kaur, Speaker Recognition Based on Cuckoo Search Algorithm, International Journal of Innovative Technology and Exploring Engineering (IJITEE), ISSN: 2278-3075, Volume-2, Issue-5, Pp. 311-313, and April 2013.

[9] Moe Moe Zaw, Ei Ei Mon, Web Document Clustering Using Cuckoo Search Clustering Algorithm based on Levy Flight, International Journal of Innovation and Applied Studies, ISSN 2028-9324, Vol.4, No.1, Pp182-188, Sep 2013.

[10] Constantine A.Balanis, Antenna Theory Analysis and Design, 2nd Edition, Pp. 310.

[11] Robert S. Elliott, Antenna Theory and Design, Revised Edition, IEEE press.

[12] Randy I. Haupt, Wiley, Antenna Array - A Computational Approach, IEEE press.

[13] Shraddha Shrivastava, Kanchan Cecil, Performance Analysis of Linear Antenna Array Using Genetic Algorithm, International Journal of Engineering and Innovative Technology (IJEIT), Volume 2, Issue 5, Pp. 84-88, November 2012.

[14] M. Khodier and M. Al-Aqeel, Linear and Circular Array Optimization: A Study of Particle Swarm Intelligence, Progress In Electromagnetics Research B, Vol. 15, 347–373, 2009.